\documentclass[twocolumn,pra,aps,superscriptaddress]{revtex4-1} 
\usepackage{mathptmx}
\usepackage{subfigure}
\usepackage{psfrag,graphicx}
\usepackage{dcolumn}
\usepackage{amsmath,amssymb}
\usepackage{bm}
\usepackage{color}
\usepackage{latexsym}
\usepackage{epstopdf}
\definecolor{mygrey}{gray}{0.35}
\definecolor{myblue}{rgb}{0.2,0.2,0.8}
\definecolor{myzard}{cmyk}{0,0,0.05,0}
\definecolor{mywhite}{rgb}{1,1,1}
\definecolor{myred}{rgb}{1,0.,0.3}
\usepackage[colorlinks=true,citecolor=myblue,linkcolor=myred]{hyperref}

\usepackage[english]{babel}

\usepackage{amsfonts}
\usepackage{natbib}
\usepackage{appendix}
\usepackage{bbold}
\usepackage{placeins}

\def\ii{{\boldsymbol{i}}}
\def\jj{{\boldsymbol{j}}}
\def\eX{{\boldsymbol{e}_x}}
\def\eY{{\boldsymbol{e}_y}}
\def\eXY{{\boldsymbol{e}_{x,y}}}

\begin{document}

\title{Topological phases of shaken quantum Ising lattices}
\author{Samuel Fern\'andez-Lorenzo}
\email{S.Fernandez-Lorenzo@sussex.ac.uk} 
\affiliation{Department of Physics and Astronomy, University of Sussex, Falmer, Brighton BN19QH, United Kingdom}
\author{Juan Jos\'e Garc{\'\i}a-Ripoll}
\affiliation{Instituto de F{\'\i}sica Fundamental IFF-CSIC, Calle Serrano 113b, Madrid E-28039, Spain}
\author{Diego Porras}
\email{D.Porras@sussex.ac.uk}
\affiliation{Department of Physics and Astronomy, University of Sussex, Falmer, Brighton BN19QH, United Kingdom}
\begin{abstract}
The quantum compass model consists of a two-dimensional square spin lattice where the orientation of the spin-spin interactions depends on the spatial direction of the bonds. It has remarkable symmetry properties and the ground state shows topological degeneracy. The implementation of the quantum compass model in quantum simulation setups like ultracold atoms and trapped ions is far from trivial, since spin interactions in those sytems typically are independent of the spatial direction.
Ising spin interactions, on the contrary, can be induced and controlled in atomic setups with state-of-the art experimental techniques.
In this work, we show how the quantum compass model on a rectangular lattice can be simulated by the use of the photon-assisted tunneling induced by periodic drivings on a quantum Ising spin model. 
We describe a procedure to adiabatically prepare one of the doubly-degenerate ground states of this model by adiabatically ramping down a transverse magnetic field, with surprising differences depending on the parity of the lattice size. 
Exact diagonalizations confirm the validity of this approach for small lattices. Specific implementations of this scheme are presented with ultracold atoms in optical lattices in the Mott insulator regime, as well as with Rydberg atoms.
\end{abstract}
\maketitle 
%


\section{Introduction}

In the pursuit of the quantum computer, the problem of decoherence arises as the main obstacle to preserve coherent linear superpositions as to take advantage of the computational power they can provide us with. In principle, quantum error correction codes offer a solution to achieve a fault-tolerant quantum computation \cite{Shor95pra}. An alternative route consists of using topologically protected Hilbert spaces \cite{Kitaev03ap,Ioffe12rpp}. In this context, a two dimensional quantum compass model on a square lattice was proposed by Dou\c{c}ot \textit{et al}. \cite{Doucot05prb} as a simple model to implement a protected qubit. Generally speaking, `compass models' refers to a broad type of lattice Hamiltonians in which the couplings between sites depend on the orientation of the bonds in the lattice. A thorough review of these Hamiltonians and their properties can be found in Ref. \cite{Nussinov15rmp}. 

The quantum compass model was originally introduced in 1982 as a toy model to gain insight in the context of Mott insulating transition metal compounds, for which one finds anisotropy of the exchange for different pairs of ions. The name `compass model' arises by analogy with the dipolar coupling in a classical model of magnetic needles arranged in a lattice \cite{Kugel82}. The 2D-version of this model on a $n\times m$ lattice is defined by the following  spin Hamiltonian ($S=1/2$),

\begin{equation}
	H_{\rm C}=-J_x\sum_{\jj} \sigma^x_{\jj}\sigma^x_{\jj + \eX}-J_y \sum_{\jj} \sigma^y_{\jj}\sigma^y_{\jj + \eY}, \label{compass}
\end{equation}
where $\sigma^{x,y}_{\jj}$ are the usual Pauli matrices and $\jj$ runs over the lattice sites. We shall assume free boundary conditions in the following. 
We can choose ferromagnetic couplings ($J_x, J_y > 0$) without lose of generality, since ferromagnetic and antiferromagnetic quantum compass models are related by unitary transformations. One observes in Hamiltonian \eqref{compass} that there are two competing tendencies owing to two types of Ising-like interactions: 
bonds along the \textit{y} axis induce spin alignment along \textit{y} $(\langle\sigma^y_{\jj}\rangle \neq 0 )$, while bonds along the \textit{x} axis 
induce spin alignment along \textit{x} $(\langle\sigma^x_{\jj}\rangle \neq 0)$; the resulting ground state is therefore a highly entangled state without an obvious order parameter. 
Many recent numerical studies have examined the quantum phase transition of the anisotropic model on a square lattice through the isotropic point ($J_x=J_y$), pointing to the existence of a first order quantum phase transition \cite{Dorier05prb,Fang07prb,Orus09prl}. 
It has been theoretically shown that this model arises as an effective description of the low-energy physics in systems of magnetically frustrated Josephson junction arrays \cite{Doucot05prb,Ioffe12rpp}, and experiments have shown signatures of the physics of this model in few qubit setups \cite{Gladchenko09natphys}.

\begin{figure}[h!]
  \includegraphics[width=0.45\textwidth]{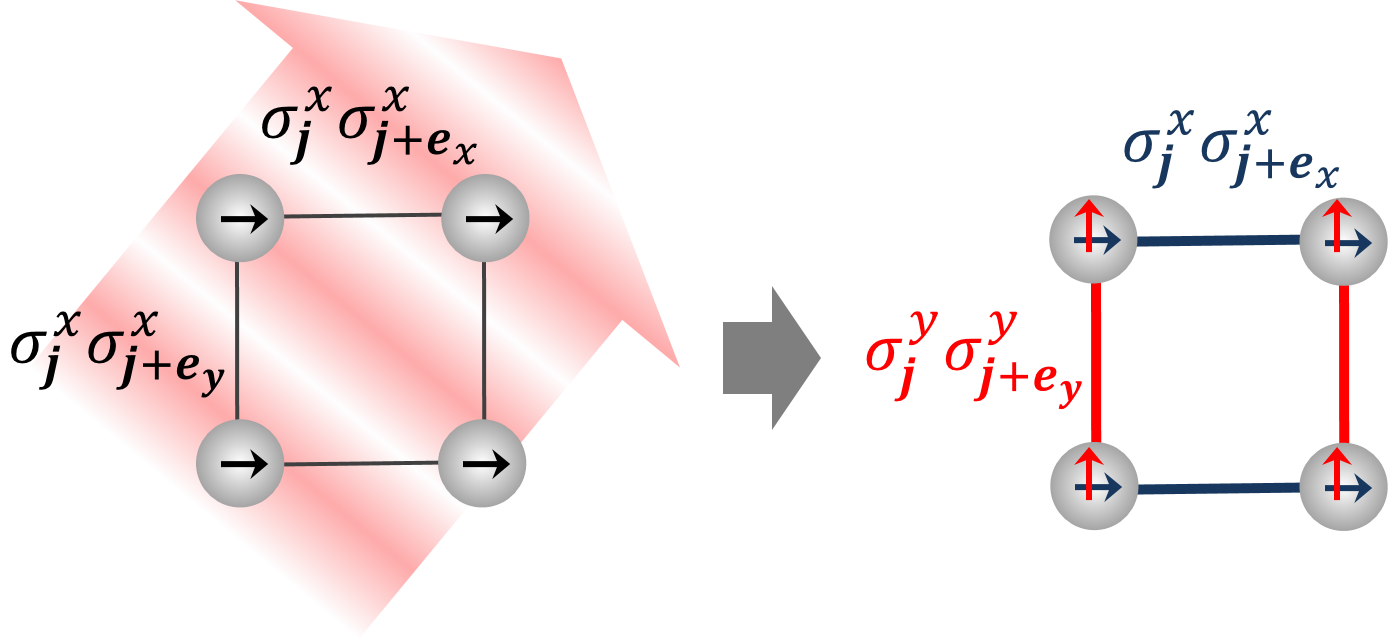}
  \caption{General scheme: Our model assumes an initial quantum Ising lattice as a starting point. Periodic drivings with a site-dependent phase allow us to dress the original Ising interaction to obtain the quantum compass model, in which sites interact through spin-components depending on the orientation of the bond.}
  \label{fig:fig0} 
\end{figure}

The physical implementation of the quantum compass model (\ref{compass}) in atomic systems would represent a significant breakthrough in research on topologically protected qubits. 
However, experimental tecniques for analogue quantum simulation \cite{Cirac12natphys} typically provide us with effective spin-spin interactions that are independent of the spatial direction of the bonds. 
For instance, Ising interactions, with couplings of the form $\sigma_\ii^x \sigma^x_\jj$ along every spatial direction, can be readily induced and controlled in systems like trapped ions
\cite{Porras04prl,Friedenauer08natphys,Schneider12rpp}, 
ultracold atoms in optical lattices \cite{Duan03prl,Garcia-Ripoll03njp,Simon11nat,Struck13natphys} 
and Rydberg atoms \cite{Weimer10natphys,Gaetan09natphy,Viteau11prl}.
In this work, we surpass this limitation, by showing that the quantum compass interactions can be implemented by dressing Ising spin-spin interactions with periodic driving fields. 
The basic idea of our work is the dressing of Ising interactions by the \textit{photon-assisted tunneling} induced by periodic drivings with a site-dependent phase over a square spin lattice . 
A judicious choice of the site-dependence of the driving phase leads to the spatial dependence of interactions in the quantum compass model. (see Fig. \ref{fig:fig0}).
The necessary periodic drivings can be implemented with running spin-dependent optical potentials such as those demonstrated in \cite{Mandel03prl}. 
Our ideas have a direct application in the implementation of topological models with ultracold atoms and Rydberg atoms with state-of-the-art techniques, since basically we only request an additional spin-dependent moving lattice to the trapping optical lattice potential. 
Indeed, periodically driven atomic lattices with site dependent phases have brought a lot of attention in recent years, as they can be used in the simulation of synthetic gauge fields \cite{RefWorks:7,RefWorks:34,Hauke12prl,Struck12prl,Aidelsburger13prl,Ketterle13prl,Goldman14rpp,Bermudez15arX,Aidelsburger15natphys}. The dressing of one-dimensional quantum Ising systems by periodic drivings with a gradient in intensity has been considered recently in \cite{Almeida13jpb}.

This article is organized as follows. In Sec.II we review some general symmetry properties of the quantum compass model that will be necessary to understand the adiabatic preparation of the ground state. 
In Sec.III the photon-assisted tunneling scheme is applied to engineer the quantum compass model using a 2D Ising model together with a convenient periodic driving. 
In Sec.IV we discuss an adiabatic preparation of the ground state of this model; the corresponding implementation of this procedure using ultracold atoms or Rydberg atoms in optical lattices is presented in Sec.V. Finally, Sec.VI summarizes the main results obtained in this work.


\section{Symmetry properties of the compass model}
Let us now look over the main symmetry properties of the quantum compass model that we shall use throughout this work \cite{Doucot05prb}. Generally speaking, a symmetry operation represented by certain operator $P$ commuting with the Hamiltonian $H$, may lead to the presence of degenerate states. This means that $P$ acting on a state $|\psi\rangle$ results in another state $|\varphi\rangle$ with the same energy, although this argument fails in the case we get the same state, $|\psi\rangle =|\varphi\rangle $. Nonetheless, if we find another symmetry operation $Q$ such that $[P,Q]\neq 0$ for any state $\left|\psi\right\rangle$, one can be sure that all states are at least doubly degenerate. To see how this works, let us suppose that acting on the state $\left|\psi\right\rangle$ with $P$ and $Q$ we obtain the same state $\left|\psi\right\rangle$. This implies that $PQ|\psi\rangle=QP|\psi\rangle=|\psi\rangle$, hence $[P,Q] = 0$, which is contrary to our initial statement. We have therefore proved that the resulting states are necessarily different. Leaving aside accidental degeneracy, another condition must be imposed if we wish to get doubly degenerate states avoiding further degeneracy. This condition turns out to be $[P^2,Q]=[P,Q^2]=0$ \cite{Doucot05prb}. The reason for this is that if one starts with an eigenstate of $Q(P)$ and then acting on it with $P(Q)$, the resulting state has to be  different from the original one as we just proved, but acting again on this state with $P(Q)$, on should come back to the initial state. Having two sets of non-commuting operators $\{P_i\}$ and $\{Q_j\}$, the previous conditions are generalized as $[P_i,Q_j]\neq 0 $ $\forall i,j $ and $[P_iP_j,Q_k]=[P_i,Q_jQ_k]=0$ $\forall i,j,k$ \cite{Doucot05prb}. 

The Hamiltonian \eqref{compass} has two sets of discrete symmetries satisfying the above conditions, namely

\begin{align}
P_{j_y}&=\prod^n_{j_x}{\sigma^y_{(j_x,j_y)}} \quad j_y=1,2,\ldots,m \label{Prow}\\
Q_{j_x}&=\prod^m_{j_y}{\sigma^x_{(j_x,j_y)}} \quad j_x=1,2,\ldots,n
\end{align}
i.e, each $P_{j_y}$ is the row product of $\sigma^y$ in that row while $Q_{j_x}$  is the column product of $\sigma^x$ in the column. Physically, $Q_{j_x}$ corresponds to a rotation by an angle $\pi$ around the \textit{x} axis of all the spins of the column labeled by $j_x$, while $P_{j_y}$ corresponds to a rotation by an angle $\pi$ around the \textit{y} axis of all the spins of the row labeled by $j_y$. In particular, they satisfy $\{P_{j_y},Q_{j_x}\}=0$ $\forall j_x,j_y$ and $[P_iP_j,Q_k]=[P_i,Q_jQ_k]=0$ $\forall i,j,k$. Thus, aside from accidental degeneracies,  one expects every state to be doubly and only doubly degenerate. We shall then assume that the ground state of the quantum compass model is effectively a two level system satisfying the conditions of a protected qubit. Local noise acting on a single lattice site may not commute with the symmetry operators $P_{j_y}$ and $Q_{j_x}$ corresponding to the row and column of such site, but the remaining symmetries ensures the system to remain doubly degenerate. This is the case unless at least $\min(m,n)$ of these perturbations act simultaneously over the whole lattice.
 The set of integrals of motion $P_{j_y}$ (or alternatively $Q_{j_x}$)  can be used to distinguish between the two degenerate states of the ground state since they have different quantum numbers, namely either $(p_1,\ldots,p_m)=(1,\ldots,1)$ or $(p_1,\ldots,p_m)=(-1,\ldots,-1)$. This result was proved in Ref.\cite{RefWorks:36} for a square lattice, and it can be straightforwardly generalized for a rectangular lattice. Heuristically, one may expect this result by exploring the trivial case for which $J_x=0$. In such a case the model is simplified to a set of Ising columns with ferromagnetic coupling. For simplicity, taking the square lattice $lxl$, the ground state consists of $2^l$ states defined by $m_{j_y,1}=\ldots =m_{j_y,l}=\pm1$ with $j_y=1,\ldots,l$, where $m_\jj$ is the eigenvalue of $\sigma^y_\jj$. As all the rows are identical, all the $p_{j_y}'s$ are equal as well for all $j_y$, thus we conclude the value of $p_{j_y}$ is either $+1$ or $-1$ for every row.  \\
\section{Photon-assisted tunneling}
In this section we shall use the photon-assisted tunneling toolbox \cite{RefWorks:34} to implement the quantum compass model. To understand how to achieve this goal, let us first have a look at the quantum Ising model,

\begin{equation}
	H_{0}=-J_x\sum_{\jj} \sigma^x_{\jj}\sigma^x_{\jj + \eX}-J_y \sum_{\jj} \sigma^x_{\jj}\sigma^x_{\jj + \eY}. \label{originalHam}
\end{equation}
Expressing the spin operators in terms of the raising and lowering operators, i.e. $\sigma^x=(\sigma^+ + \sigma^-)$ and $\sigma^y=-i(\sigma^+ - \sigma^-)$, Eq. \eqref{originalHam} turns out to be,
\begin{multline}
	H_{0}=-J_x \sum_{\jj}\left( \sigma^+_{\jj}\sigma^+_{\jj + \eX}+\sigma^+_{\jj}\sigma^-_{\jj + \eX}\right) \\
	-J_y \sum_{\jj}\left( \sigma^+_{\jj}\sigma^+_{\jj + \eY} +\sigma^+_{\jj}\sigma^-_{\jj + \eY}\right)
	+ {\rm H.c.} \label{originalHam2}
\end{multline}
In contrast, rewriting the quantum compass model \eqref{compass} in a similar way, we obtain a slightly different expression
	{\rm H.c.} \label{compass2}
\begin{multline}
	H_{C}=-\sum_{\jj}\left(J^{++}_x\sigma^+_{\jj}\sigma^+_{\jj + \eX}+J^{+-}_x\sigma^+_{\jj}\sigma^-_{\jj + \eX}\right) \\
	-\sum_{\jj}\left( -J^{++}_y\sigma^+_{\jj}\sigma^+_{\jj + \eY} +J^{+-}_y\sigma^+_{\jj}\sigma^-_{\jj + \eY}\right) + {\rm H.c.} ,\label{compass2}
\end{multline}
where we have defined $J^{++}_{x,y}=J^{+-}_{x,y}=J_{x,y}$. By the use of the photon assisted tunneling in the original Hamiltonian \eqref{originalHam}, we aim for finding a set of effective coupling constants such that Eq.\eqref{originalHam2} equals Eq. \eqref{compass2}, which in turn means
\begin{align}
(J^{++}_{x})_{\rm eff}&=(J^{+-}_{x})_{\rm eff} \label{idea}\\ 
(J^{++}_{y})_{\rm eff}&=-(J^{+-}_{y})_{\rm eff} \nonumber .
\end{align}
Now that our goal is clear, let us use the ingredients of the photon-assisted tunneling toolbox to find how to achieve the conditions \eqref{idea}. In doing so, the Hamiltonian of the system can be written as,
\begin{equation}
H = H_0+H_d(\tau), \label{patGen}
\end{equation}
where we define,
\begin{equation}
H_{\rm d}(\tau ) = \sum_{\jj} \frac{\Omega_{\jj}}{2} \sigma^z_{\jj} + 
\sum_{\jj}\frac{\eta}{2}\omega_d\cos(\omega_{\rm d}\tau+\phi_{\jj})\sigma^z_{\jj} . \label{pat}
\end{equation}

The second term in \eqref{pat} represents a periodic energy driving of the qubit, while $\Omega_\jj$ is typically chosen such that the first term in Eq.\eqref{pat} represents a gradient of the individual frequencies \cite{RefWorks:7}, i.e., $\Omega_j=\Omega_0+\Delta\Omega\cdot j$, although it will be taken as a constant for the purpose of this work, $\Omega_\jj=\Omega$. These two elements are all that we need to take advantage of the photon-assisted tunneling toolbox \cite{RefWorks:34}. Notice that there is a freedom in choosing the spatial dependence in $\phi_\jj$, and we shall assume a linear dependence in both \textit{x} and \textit{y} axis,
\begin{equation}
	\phi_\jj=\Delta\phi_x j_x+\Delta\phi_y j_y \label{phase},
\end{equation}
where $\Delta\phi_{x,y}$ are given constants, and $j_{x,y}$ are positions in the lattice. 

Let us now express the Hamiltonian $H_0$ in the interaction picture with respect to $H_d$, namely $H_0(t)=U(t)^{\dagger}H_0 U(t)$ where $U(t)=e^{-i\int^{t}_{0}d\tau H_d(\tau)}$. In this picture, the raising and lowering operators evolve like,
\begin{equation}
\sigma^\pm_{\jj}(t)=e^{\pm i\Omega t}e^{\pm i\eta\sin{(\omega_{\rm d}t+\phi_{\jj}})}e^{\mp i\eta\sin{\phi_{\jj}}}\sigma^\pm_{\jj} . \label{evolutionOp}
\end{equation}
Notice that the last term in Eq.\eqref{evolutionOp} can be gauged away using the unitary transformation $\sigma^\pm_{\jj}\rightarrow e^{\pm i\eta\sin{\phi_{\jj}}}\sigma^\pm_{\jj} $. Hence one obtains a Hamiltonian having the same structure as Eq.\eqref{compass2}, in which we replace the bare couplings by their corresponding time-dependent dressed couplings
\begin{align}
(J^{++}_{x,y})_{\rm eff}&= J_{x,y} e^{i2\Omega\tau}e^{i\eta \sin{(\omega_d\tau+\phi_{\jj})}}e^{i\eta \sin{(\omega_d\tau+\phi_{\jj+\eXY})}}\\
(J^{+-}_{x,y})_{\rm eff}&=J_{x,y} e^{i\eta \sin{(\omega_d\tau+\phi_{\jj})}}e^{-i\eta \sin{(\omega_d\tau+\phi_{\jj+\eXY})}} .
\end{align}
To help in the analytical treatment, we make use of the Jacobi-Anger expansion,
\begin{equation}
e^{(iz\sin\phi)}=\sum^{\infty}_{n=-\infty} \mathcal{J}_n(z)e^{(in\phi )},
\end{equation}
where $\mathcal{J}_n$ are Bessel functions of the first kind, yielding 
\begin{align}
(J^{++}_{x,y})_{\rm eff} &= J_{x,y} e^{i2\Omega\tau}\sum_{s,s'} J_s(\eta) J_{s'}(\eta ) e^{is(\omega_d\tau+\phi_\jj )}e^{is'(\omega_d\tau+\phi_{\jj+\eXY} )} \label{J++} \\
(J^{+-}_{x,y})_{\rm eff} &= J_{x,y} \sum_{s,s'} J_s(\eta) J_{s'}(\eta ) e^{is(\omega_d\tau+\phi_\jj )}e^{-is'(\omega_d\tau+\phi_{\jj+\eXY} )} \label{J+-}.
\end{align}
Assuming $J_{x,y} \ll \Omega $ and tuning the driving frequency to $\omega_d=2\Omega$, the rotating-wave approximation (RWA) allows us to neglect the fast-oscillating  terms and to keep those terms fulfilling the resonance condition $s'=s+1$ in Eq.\eqref{J++} and $s'=s$ in Eq.\eqref{J+-}. Having done this, we obtain the following dressed couplings,
\begin{align}
(J^{++}_{x,y})_{\rm eff} &= J_{x,y}\mathcal{F}_{x,y}^{++}(\eta ,\Delta\phi_{x,y} )e^{-i\left(\frac{\phi_\jj+\phi_{\jj+\eXY}}{2}\right)} \\
(J^{+-}_{x,y})_{\rm eff} &= J_{x,y}\mathcal{F}_{x,y}^{+-}(\eta ,\Delta\phi_{x,y} ),
\end{align}
where we define a set of complex amplitudes,
\begin{align}
\mathcal{F}_{x,y}^{++}(\eta ,\Delta\phi_{x,y} )&= \sum_s \mathcal{J}_s(\eta )\mathcal{J}_{-s-1}(\eta )e^{-i(s+\frac{1}{2})\Delta\phi_{x,y}} \\	
\mathcal{F}_{x,y}^{+-}(\eta ,\Delta\phi_{x,y} ) &= \sum_s \mathcal{J}_s(\eta )e^{-is\Delta\phi_{x,y}}.
\end{align}
Finally, rewriting the term $\frac{\phi_\jj+\phi_{\jj+\eXY}}{2}=\phi_\jj+\frac{\Delta\phi_{x,y}}{2}$ we may use the following gauge transformation: $\sigma^+_\jj\longmapsto\sigma^+_\jj e^{i\phi_{\jj/2}}$, resulting in the effective Hamiltonian we were looking for,

\begin{multline}
	H_{\rm eff}=-J_x \sum_{\jj}\left( \mathcal{F}_x^{++}\sigma^+_{\jj}\sigma^+_{\jj + \eX}+\mathcal{F}_x^{+-}\sigma^+_{\jj}\sigma^-_{\jj + \eX}\right) \\
	-J_y \sum_{\jj}\left( \mathcal{F}_y^{++}\sigma^+_{\jj}\sigma^+_{\jj + \eY} +\mathcal{F}_y^{+-}\sigma^+_{\jj}\sigma^-_{\jj + \eY}\right) + {\rm H.c.} \label{compass3}
\end{multline}
Following the idea given in Eq.\eqref{idea}, we notice that the quantum compass model can be implemented if we manage to find a set of parameters, $(\eta,\Delta\phi_x,\Delta\phi_y)$, such that
\begin{align}
\mathcal{F}_x^{++}(\eta,\Delta\phi_x)&=\mathcal{F}_x^{+-} (\eta,\Delta\phi_x,) \label{condiciones}\\
\mathcal{F}_y^{++} (\eta,\Delta\phi_y)&= -\mathcal{F}_y^{+-}(\eta,\Delta\phi_y) .\nonumber
\end{align}

Is there any solution to the system of equations \eqref{condiciones}? We will show numerically that there are actually an infinite number of solutions. First, note that Eq. \eqref{condiciones} can be expressed as,

\begin{align}
\left|\mathcal{F}_{x,y}^{++}(\eta,\Delta\phi_{x,y})\right|&=\left|\mathcal{F}_{x,y}^{+-} (\eta,\Delta\phi_{x,y})\right| \label{cond1}\\
\arg(\mathcal{F}_x^{++} (\eta,\Delta\phi_x)) &=\arg(\mathcal{F}_x^{+-}(\eta,\Delta\phi_x)) + 2k\pi  \label{cond2} \\
\arg(\mathcal{F}_y^{++} (\eta,\Delta\phi_y))&= \arg(\mathcal{F}_y^{+-}(\eta,\Delta\phi_y))+k\pi . \label{cond3}
\end{align}

\begin{figure}[h!]
  \subfigure[]{\label{fig:fig1:a} \includegraphics[width=0.45\textwidth]{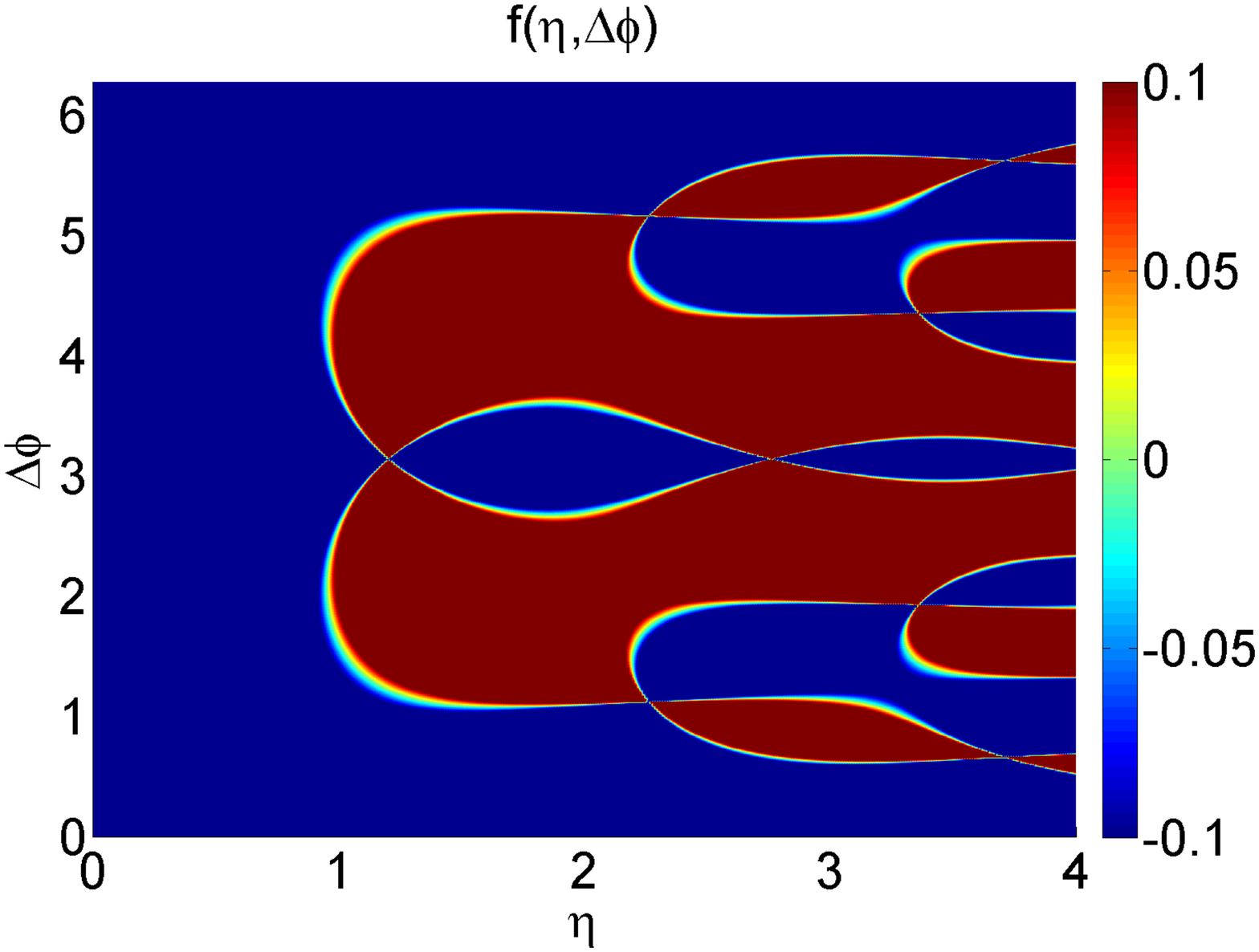}}
  \subfigure[]{\label{fig:fig1:b} \includegraphics[width=0.45\textwidth]{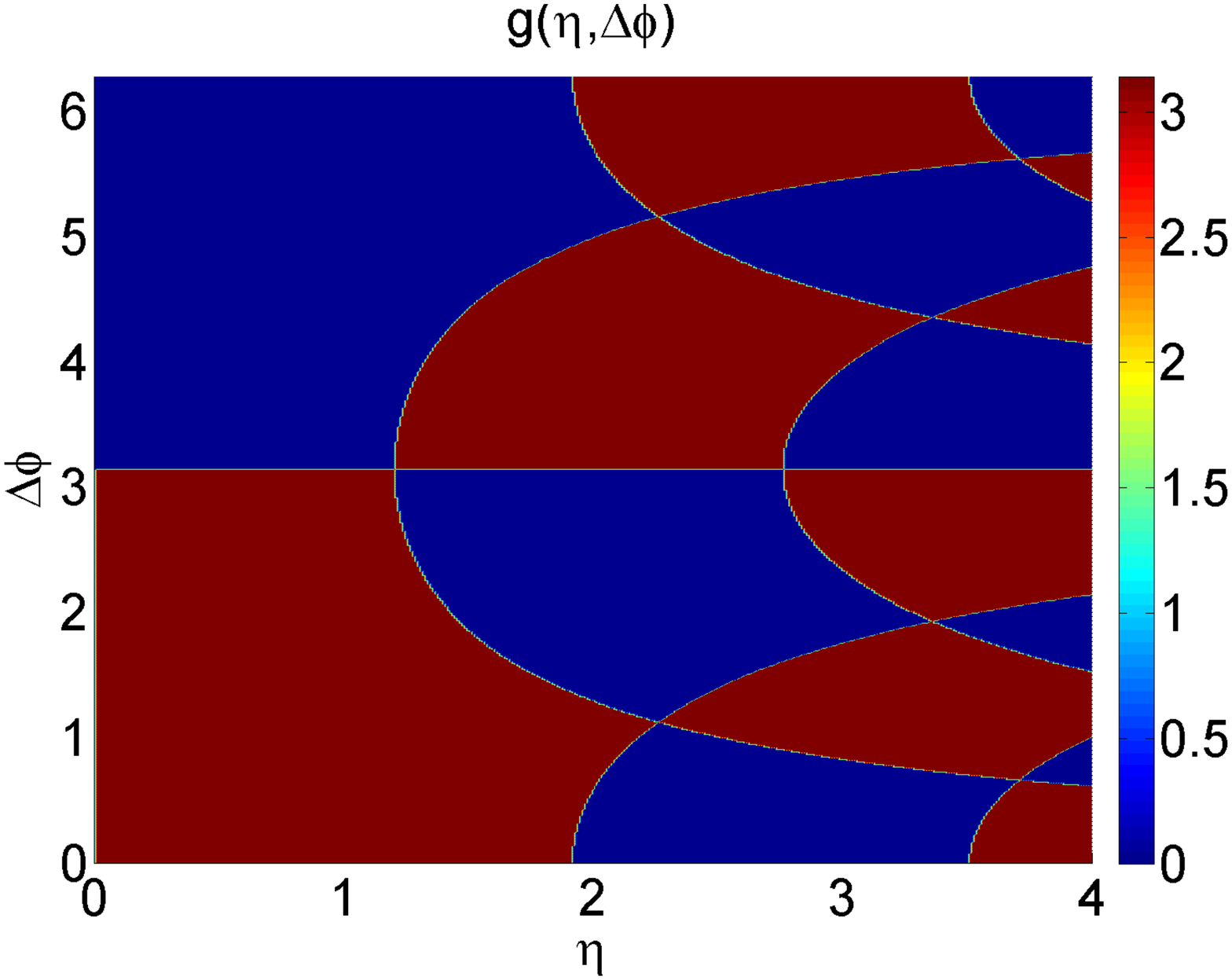} }
  \caption{In this figure we plot the functions that govern the spin-spin interactions as a function of the periodic driving parameters. \textbf{(a)} Function $f(\eta,\Delta\phi_{x,y})=\left|\mathcal{F}_{x,y}^{++}(\eta,\Delta\phi_x)\right|/\left|\mathcal{F}_{x,y}^{+-} (\eta,\Delta\phi_{x,y})\right|-1$.
  The values  $f(\eta,\Delta\phi_{x,y}) = 0$ represents the solutions to Eq.\eqref{cond1}. 
  The range of the function is limited to the interval $[-0.1,0.1]$ to show clearly the points at which the funcion is zero. \textbf{(b)} Function $g(\eta,\Delta\phi_{x,y})=\arg(\mathcal{F}_{x,y}^{++} (\eta,\Delta\phi_y))-\arg(\mathcal{F}_{x,y}^{+-}(\eta,\Delta\phi_y))$, limited to the range $[0,2\pi]$. The regions with  $g(\eta,\Delta\phi_{x,y}) = 0$ correspond to the solutions to Eq.\eqref{cond2} and Eq.\eqref{cond3}.}
  \label{fig:fig1}
\end{figure}

To solve this system of equations, let us define the functions $f(\eta,\Delta\phi_{x,y})=\left|\mathcal{F}_{x,y}^{++}(\eta,\Delta\phi_x)\right|/\left|\mathcal{F}_{x,y}^{+-} (\eta,\Delta\phi_{x,y})\right|-1$ and $g(\eta,\Delta\phi_{x,y})=\arg(\mathcal{F}_{x,y}^{++} (\eta,\Delta\phi_y))-\arg(\mathcal{F}_{x,y}^{+-}(\eta,\Delta\phi_y))$. Trivially, we see that Eq.\eqref{cond1} is equivalent to find the zeros of the function $f$, while the solutions to Eq.\eqref{cond2} and Eq.\eqref{cond3} are equivalent to the function $g$ taking the values $k\pi$ and $2k\pi$ respectively. The jellyfish-like pattern showed in figure \ref{fig:fig1:a} represents the function $f$, for which we have limited the range of the function to the interval $[-0.1,0.1]$ so that the graph clearly shows the region in which the function becomes zero. On the other hand, the function $g$ is shown in Fig.\ref{fig:fig1:b}, limited to the range $[0,2\pi]$. In looking at the graphs, it is straightforward to confirm that there are an infinite number of solutions. The isotropic point, ($J_x=J_y$) is reached when the solutions are taken symmetrically with respect to the symmetry axis of the jellyfish-like pattern; an example could be $\{\eta=1,\Delta\phi_x=4.6539,\Delta\phi_y=1.6293\}$, which gives the value $\mathcal{F}_{x}^{++}=\mathcal{F}_{y}^{+-}=0.5368$.

\section{Adiabatic passage}

Now that we have shown how the quantum compass model can be implemented using periodic drivings, we aim for preparing the ground state of this model. One way to do so is by finding an adequate adiabatic passage starting from the ground state of certain initial Hamiltonian for which the ground state is known and can be prepared, and then slowly changing this Hamiltonian until the quantum compass model is eventually reached. One possible option using a transverse magnetic field will be discussed in this section. 

Let us then consider an additional transverse magnetic field along \textit{z} in our original Hamiltonian \eqref{patGen} as follows,
\begin{equation}
	H = H_0+H_d(\tau)+ \sum_{\jj} \frac{\delta(t)}{2} \sigma^z_{\jj} \label{transField2},
\end{equation}
where $\delta$ is a time-dependent parameter that measures the strength of the field. Notice that we had already introduced a transverse field along \textit{z} in Eq.\eqref{patGen}, so the overall transverse field depends on the sum of both terms, $\epsilon/2=(\Omega+\delta)/2$. One may also express $H_0+\sum_{\jj} \delta \sigma^z_{\jj} $ in the interaction picture with respect $H_d$; it is then straightforward to check that, following the same procedure used in Sec.III and setting the same resonance condition $\omega_d=2\Omega$, we arrive at the following effective Hamiltonian,
\begin{equation}
	H'_{\rm eff}(t)=H_{\rm eff}+\sum_{\jj} \frac{\delta(t)}{2} \sigma^z_{\jj}, \label{transField}
\end{equation}
where $H_{\rm eff}$ was given in Eq.\eqref{compass3}. Therefore, ensuring the magnetic field is strong enough so that $\delta>>\Omega,\eta>>J_x,J_y$, one can safely assume that the magnetic field is the dominant term in \eqref{transField}, being the ground state $|0\rangle=\prod_{\jj}|\uparrow\rangle_{\jj}$ in terms of eigenstates of $\sigma^z_\jj$. Finally, a feasible adiabatic passage to reach the compass model consists of decreasing $\epsilon$ from $\epsilon>>\Omega,\eta$ slowly enough until reaching the compass condition $\epsilon=\Omega=\omega_d/2$ $(\delta=0)$. One expects this system to undergo a quantum phase transition as the magnetic order of the transverse magnetic field and the quantum compass model are different.

The adiabatic approximation describes, upon certain conditions, how slowly we need to vary $\delta(t)$ to ensure the system remains in the ground state through the evolution. According to this approximation, the instantaneous eigenstates of the time-dependent Hamiltonian $H(t)$ at a given time evolve continuously to the corresponding eigenstates at later times, provided that the eigenenergies do not cross and the evolution is slow enough. The intrinsic time scale used to determine what slow and fast mean is usually provided by the gaps in the spectrum. This also provides a general validity condition for adiabatic behaviour that corresponds to the probability that the final state of the system is different from the initial state \cite{RefWorks:46},
\begin{equation}
	\max_{0\leq t \leq T}\left|\frac{\langle k|\dot{H}(t)|n\rangle}{g_{nk}}\right|<< \min_{0\leq t \leq T}|g_{nk}| \label{adiabatic}
\end{equation}
where $T$ is the total evolution time, and $g_{nk}$ the energy gap between level $n$ and $k$. Note that the ground state of the transverse field Hamiltonian is unique, while the compass ground state is doubly-degenerate, thus we expect crossing of levels at the end of the adiabatic preparation; moreover, the natural question arises about what state is eventually reached. 

\begin{figure}[p]
 \subfigure[]{\label{fig:fig2:a} \includegraphics[width=0.5\textwidth]{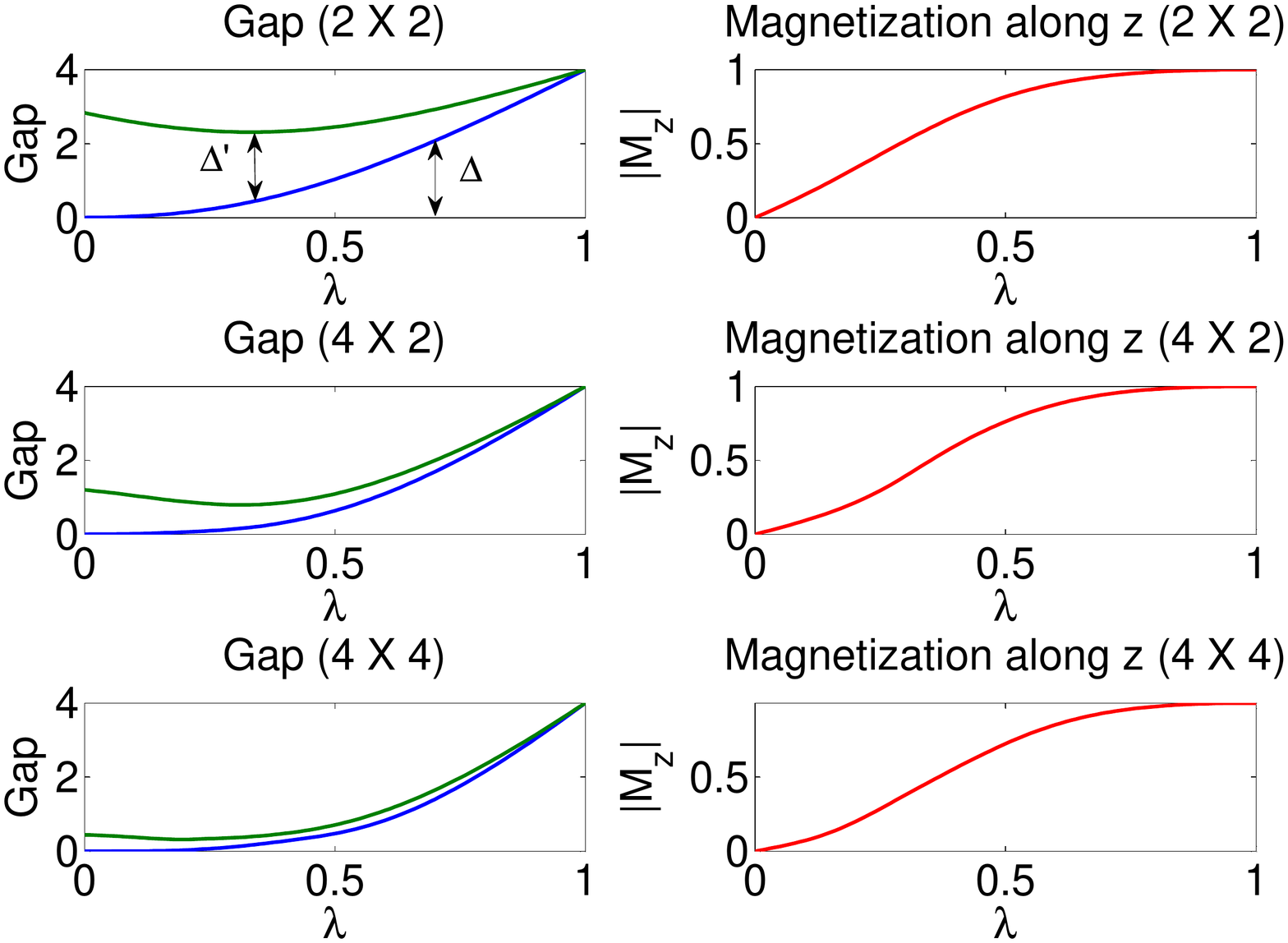}}
  \subfigure{\label{fig:fig2:b} \includegraphics[width=0.5\textwidth]{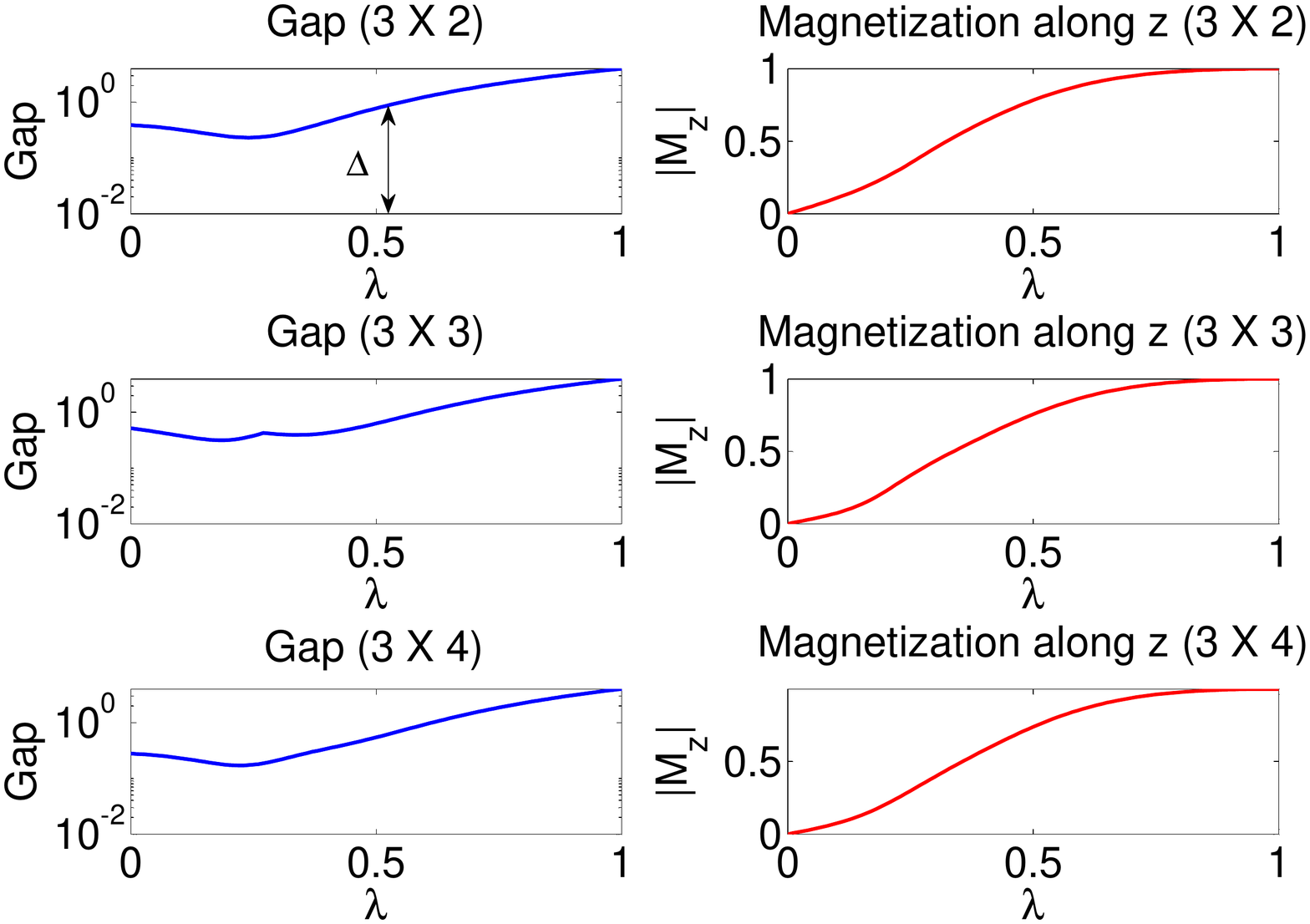} }
	\addtocounter{subfigure}{-1}
	\subfigure[]{\label{fig:fig2:c} \includegraphics[width=0.5\textwidth]{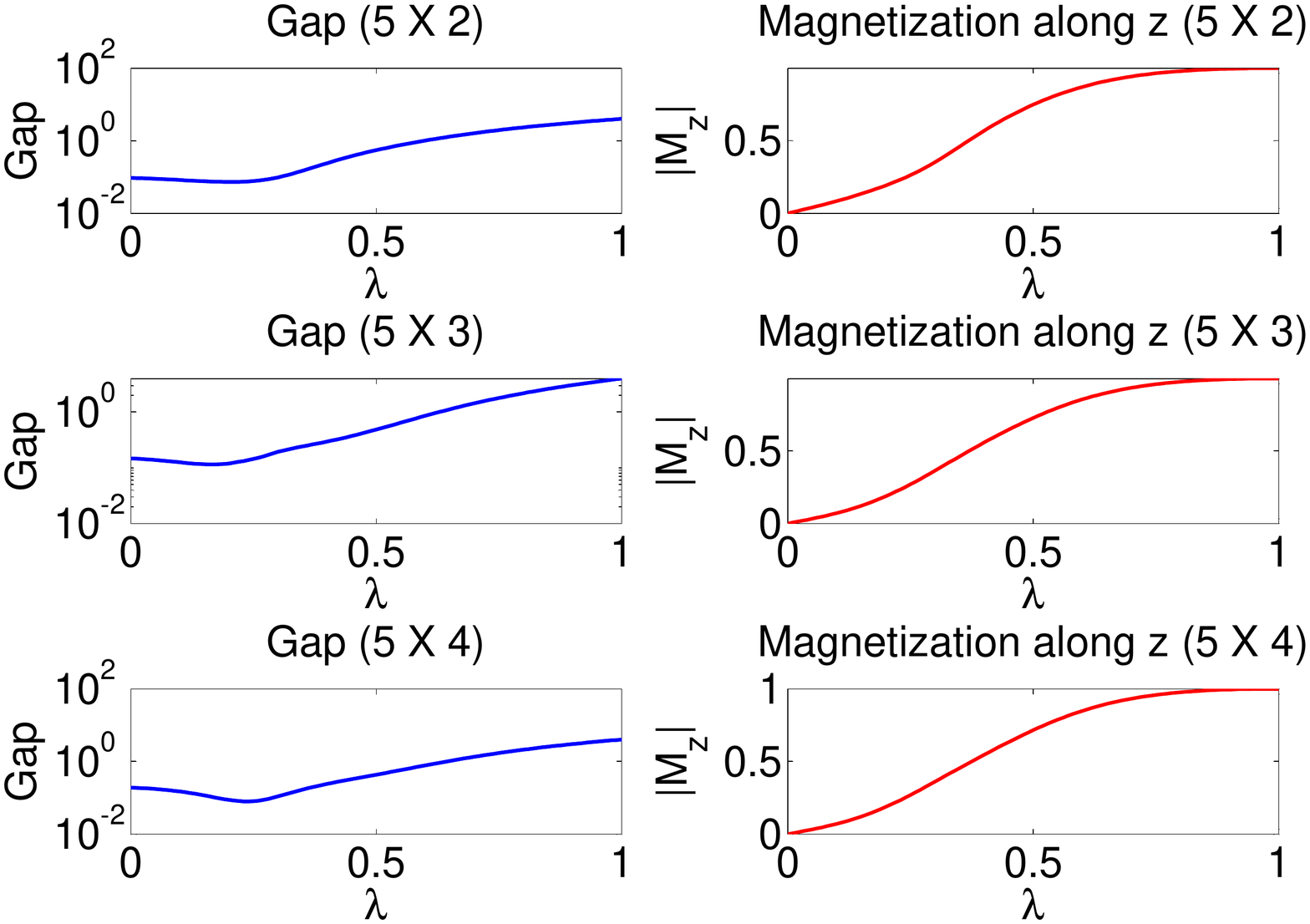}}
  \caption{Quantum phase transition from transverse field to compass model for (a) even-even lattice size and (b) odd-even/odd-odd lattice size. We show the gaps together with the absolute value of the normalized magnetization along z of the model \eqref{QPT} for different values of the lattice size. The blue and green lines corresponds, respectively, to the gap between the first and second even excited states with respect to the ground state. The gaps in Figure (b) are represented in a semi-log plot. The value $\lambda=1$ corresponds to the initial transverse field Hamiltonian and $\lambda=0$ represents the isotropic quantum compass model. }
  \label{fig:fig2}
\end{figure}

A simple possible way to overcome the issues above is by finding an integral of motion conserved through the evolution that we may use to tell the twofold ground state apart. Under this assumption, the adiabatic approximation would still be valid if there is no crossing of levels with the same conserved quantum number. As it was shown above, the set of integrals of motion $P_{j_y}(Q_{j_x})$ can be used to distinguish between the two degenerate states, since they have different quantum numbers, namely either $(p_1,\ldots,p_M)=(1,\ldots,1)$ or $(p_1,\ldots,p_M)=(-1,\ldots,-1)$ (similarly for the quantum numbers $(q_1,\ldots,q_N)$) ; however these are not good quantum numbers when the magnetic field term is included in the Hamiltonian. The operator $Z\equiv\prod_{\jj}\sigma^z_{\jj}$ is fortunately an integral of motion of the whole Hamiltonian \eqref{transField} as it is straightforward to check, and so it can be used in principle to determine the actual ground state reached at the end of the adiabatic passage. Surprisingly, the parity operator $Z$ can distinguish between the twofold ground state only in those cases such that we have an odd-odd or odd-even lattice size. In Appendix I, it is proved that for both odd-odd and odd-even cases the eigenvalues of $Z$ are either $z=+1$ or $z=-1$, while in the even-even case the eigenvalue is always $z=+1$. Hence we expect two different kinds of behaviour in the adiabatic passage depending on the parity of the lattice size. In the first case, the two ground states of the compass model are not connected, thereby the only relevant gap to be considered is between the ground and first even excited state. Recall that the ground state of the transverse field Hamiltonian has eigenvalue $z=+1$, so we expect to prepare the ground state of the compass model corresponding to this eigenvalue; however we would be unable to prepare the ground state corresponding to the eigenvalue $z=-1$ using this approach. In the second case, with an even-even lattice size, any superposition of the ground states are eigenstates of the operator $Z$ with eigenvalue $z=+1$, and for that reason, in this case we could only assume that the final state is a superposition state given by the specific adiabatic evolution performed on the system. The adibaticity of this case would be given by the gap between the first and and second even excited states. 

Exact diagonalizations for small systems (up to a $5x4$ lattice) were performed to observe the evolution of the relevant gaps involved in the quantum phase transition so as to examine the validity of the adiabatic approximation. Additionally, the magnetization along $z$ was found to be an order parameter of the quantum phase transition, where the ground state of the transverse magnetic field is magnetically ordered, $M_z\neq 0$, and the compass ground state is disordered, $M_z=0$. In order to simplify the analysis, the following parametrization was introduced,
\begin{align}
	H &=(1-\lambda) H_{C}+\lambda H_{m} \label{QPT}\\
	H_{C}&=-\sum_{\jj} \sigma^x_{\jj}\sigma^x_{\jj + \eX}- \sum_{\jj} \sigma^y_{\jj}\sigma^y_{\jj + \eY} \nonumber \\
	H_{m} &= \sum_{\jj}\sigma^z_{\jj} \nonumber ,
	\label{adiabatic}
\end{align}
therefore $\lambda=1$ corresponds to the initial transverse field and $\lambda=0$ represents the isotropic quantum compass model. 

Keeping in mind the results for the integral of motion $Z$, the relevant gap in an odd-odd and odd-even lattice is the one given by the first even excited state of the system, $\Delta$. In contrast, this gap goes to zero in an even-even lattice, so the relevant gap in this case, $\Delta'$, is given by the energy difference with the second even excited state. Note that those levels with odd parity ($z=-1$) do not play any role in any case as the Hamiltonian does not connect states with different parities. Fig.\ref{fig:fig2} shows the evolution of these gaps for different sizes together with the magnetization $|M_z|$ as the parameter $\lambda$ changes from $\lambda=1$ to $\lambda=0$. For the odd-odd and odd-even cases we confirm that there exists a finite gap along the adiabatic passage between the ground state and the first even excited state that narrows as the lattice size becomes larger. As expected, this gap goes to zero in the even-even case when the two degenerate even ground states collide in the compass model as shown in Fig.\ref{fig:fig2:b}. Therefore, the relevant gap when we approach $\lambda=0$ is given by the energy difference between the first and second even excited states $\Delta'$. In concluasion, these results show that the even ground state of the compass model can be prepared using this adiabatic passage in the odd-odd and odd-even cases , whereas for the even-even case we expect the system to be in a superposition of the ground states given by the specific evolution of $\delta(t)$.


\section{Physical implementations}
In this section we show how our ideas can be implemented using specific atomic experimental setups. 
One possible implementation of the quantum compass model using Josephson junctions arrays was proposed by Dou\c{c}ot.et.al \cite{Doucot05prb}. A close related model was actually implemented in a proof-of-principle experiment using superconducting nanocircuits.

Atomic systems present many advantages for quantum state preparation and measurement. Thus, a scalable and efficient implementation of the quantum compass model in atomic experimental setups would be very useful. 
Furthermore, implementing a controllable longitudinal magnetic field like in Eq. (\ref{adiabatic}) may allow experimentalist to adiabatically create the topologically degenerate ground state.
Periodic drivings like those required for our proposal can be implemented in atomic systems by means of lasers, with a site dependent phase that corresponds to the laser optical phase. Ions trapped in two-dimensional arrays of microtraps or Coulomb crystals could be considered here, because optical forces can be used to induce Ising interactions \cite{Schneider12rpp,Porras04prl}. However, the dipolar decay of trapped ion spin-spin interactions would lead to long-range quantum compass models, with properties that may depart from the original Hamiltonian (\ref{compass}). In the following we focus on proposals for atomic setups that may provide us with short-range Ising interactions, namely, neutral bosonic ultracold atoms and Rydberg atoms in optical lattices.

\subsection{Ultracold bosons in optical lattices}
We need first to understand how the following effective quantum Ising Hamiltonian can be implemented with ultracold bosons. For this we rely on a quantum simulation proposal which relies on the internal state of atoms that are frozen in a Mott insulator state in an optical lattice \cite{Duan03prl,Garcia-Ripoll03njp}. Under appropriate circumstances, hopping can only be a virtual process that enables superexchange interactions, as explained in \cite{Garcia-Ripoll03njp} through perturbative calculations. These interactions have been demonstrated experimentally, both in superlattices \cite{Trotzky08sci,Nascimbene12prl}, as well as in longer tubes with a few quasiparticle excitations \cite{Fukuhara13nat,Fukuhara13natphys}. 

We consider an optical lattice in the Mott-insulator regime with one atom per site (unity filling) and each atom having two accessible internal states, playing the role of a pseudo-spin $S=1/2$.
The atoms may be formally identified with two types of bosons, $'\uparrow'$, and $'\downarrow'$, and one may denote the bosonic operators $a_{\ii}$ and $b_{\ii}$ as the destruction operators of each internal state at the site $\ii$. Such a system is well described with the Bose-Hubbard Hamiltonian when the energies involved are small enough so that the second Bloch band never gets populated,  
\begin{align}
	H &= \sum_{\langle \ii,\jj\rangle} 
	H^{\rm hop}_{\ii,\jj}+\sum_\jj H^{{\rm int}}_{\jj}  \\
	H^{\rm hop}_{\ii,\jj}&=-J_a(a_\ii^{\dag}a_\jj+h.c.)+J_b(b_\ii^{\dag}b_\jj+h.c.) \nonumber \\
	H^{{\rm int}}_{\jj}&=  \frac{1}{2}U_{aa}a_\jj^{\dag}a_\jj^{\dag}a_\jj a_\jj+\frac{1}{2}U_{bb}b_\jj^{\dag}b_\jj^{\dag}b_\jj b_\jj+U_{ab}a_\jj^{\dag}b_\jj^{\dag}b_\jj a_\jj \nonumber,
\end{align}
The Hamiltonians $H^{{\rm hop}}$ and $H^{{\rm int}}$ represent, respectively, the probability of atoms hopping to neighbouring sites and their effective on-site interaction.  
Assuming we are in the Mott-insulator regime, $J \ll U$, the hopping Hamiltonian can be considered as a small perturbation with respect to all other terms. We project our problem into the subspace of single atomic occupation and use quasi-degenerate second-order perturbation theory to obtain an effective spin Hamiltonian. The corresponding operators of the effective spin system are 
$\sigma^+_{\jj} = a^\dagger_{\jj} b_{\jj},$ 
$\sigma^-_{\jj} = a_{\jj} b^\dagger_{\jj}$ and 
$\sigma^z_{\jj} = a^\dagger_\jj a_\jj - b^\dagger_\jj b_\jj$. The effective spin Hamiltonian reads
\begin{equation}
	H_{\rm S} 
	= \sum_{\langle \ii,\jj\rangle} \left(
	\lambda^z \sigma^z_{\ii} \sigma^z_{\jj} +
	\lambda^\perp (\sigma^x_{\ii} \sigma^x_{\jj} +          \sigma^y_{\ii} \sigma^y_{\jj} ) \right)
	+ \sum_{\ii} h^z \sigma^z_{\jj}  
\end{equation}
with constants given by \cite{Garcia-Ripoll03njp}

\begin{eqnarray}
\lambda^z &=&
\frac{J_a^2+J_b^2}{2U_{ab}} - \frac{J_a^2}{U_{aa}} - \frac{J_b^2}{U_{bb}}, 
\quad
\lambda^\perp = - 2 \frac{J_a J_b}{U_{ab}} ,
\nonumber \\
h^z &=& \frac{J_a^2}{U_{aa}}-\frac{J_b^2}{U_{bb}}.
\end{eqnarray}
For details on the range of validity and the derivation of those equations we refer the reader to Ref. \cite{Garcia-Ripoll03njp}. We notice that if the term $\lambda^\perp = 0$, then we can use the remaining Ising interaction as a starting point to derive the quantum compass model. 
To be in a regime such $\lambda^\perp = 0$, one could suppress the tunneling of $b$-atoms by choosing a spin-dependent lattice with $J_b \ll J_a$. If we neglect $J_b$ we get
\begin{equation}
H_{\rm S} = - J \sum_{\langle \ii,\jj\rangle} \sigma^z_{\ii} \sigma^z_{\jj} 
+ \sum_{\ii} h^z \sigma^z_{\jj},  
\label{HS}
\end{equation}
with $J = J_a^2 \left(1/U_{aa} - 1/(2 U_{ab}) \right)$, and $h^z = -J_a^2/U_{aa}$.

To implement periodic drivings, we consider additional lasers inducing Raman transitions between levels $\uparrow$ and $\downarrow$. Here, the spatial dependence of the phase will appear naturally, since the optical phase of the lasers vary linearly from site to site. We need to implement the driving and longitudinal field in Eq. (\ref{pat}). 
Note that, since we obtained an Ising interaction in the $z$-basis, the longitudinal fields required in Eq. (\ref{pat}) must be expressed in terms of $\sigma^x$ operators. A constant  field is implemented by a two-photon Raman transition or microwave field inducing transitions between the two atomic levels,
\begin{equation}
H_{\rm mw} = \frac{\Omega}{2} \sum_{\ii} \sigma^x_{\ii}.
\label{Hmw}
\end{equation}
The periodic driving fields are then implemented by a running-wave potential induced by pairs of lasers with effective wavevector $\Delta {\bf k}$, and relative detuning within each pair $\omega_{\rm d}$,
\begin{equation}
H_{\rm las}(t) = 
\frac{\Omega_{\rm las}}{2} 
\sum_{\ii} \cos(\Delta {\bf k} \ {\bf r}_\ii - \omega_{\rm d} t) \sigma^x_\ii.
\label{Hlas}
\end{equation}
Note that time-dependent optical lattice potentials have been implemented for spin-dependent transport of ultracold bosons in optical lattices (see for example \cite{Mandel03prl}). 
We notice that the laser optical phase is translated into a site-dependent optical phase with a linear dependence on the site position \cite{RefWorks:7,RefWorks:34},
\begin{equation}
\phi_\ii = d_0 (\Delta {\bf k} \  \ii),
\end{equation}
where we have used that ${\bf r}_\ii = d_0 \ii$, with $d_0$ the distance between sites in the lattice. After implementing a rotation of the spin basis ($\bar{\sigma}^{z} = \sigma^{x}$, $\bar{\sigma}^x = - \sigma^z$ ) in Hamiltonians (\ref{HS}, \ref{Hmw}, \ref{Hlas}), we obtain the driving term in Eq. (\ref{pat}), with $\eta \omega_{\rm d} = \Omega_{\rm las}$, and the phase gradient $\Delta \phi_{x,y} = (\Delta {\bf k})_{x,y} d_0$. Our final Hamiltonian is
\begin{eqnarray}
H &=& H_{\rm S} + H_{\rm mw} + H_{\rm las}(t) = \nonumber \\
&-& J \sum_{\langle \ii,\jj\rangle} \bar{\sigma}^x_{\ii} \bar{\sigma}^x_{\jj} 
+ \sum_{\ii} h^x \bar{\sigma}^x_{\jj} + 
\frac{\Omega}{2} \sum_{\ii} \bar{\sigma^z}_{\ii} + \nonumber \\
&& 
\hspace{2cm} \frac{\eta \omega_{\rm d}}{2} \sum_{\ii} \cos(\Delta {\bf k} \ {\bf r}_\ii - \omega_{\rm d} t) \bar{\sigma}^z_\ii .
\end{eqnarray}
We get our target Hamiltonian plus an additional magnetic field term, 
$h^x \sum_{\ii} \bar{\sigma}^x_\ii = h^x \sum_\ii (\bar{\sigma}^+_\ii + \bar{\sigma}^-_\ii)$. Under the resonance condition $\omega_{\rm d} = 2 \Omega$ raising and lowering operators rotate with frequency $\Omega$ 
(see Eq. (\ref{evolutionOp})). Thus, the $h^x$ term can be neglected in a rotating wave approximation in the limit $\Omega \gg J, h^x$.


\subsection{Rydberg atoms in optical lattices}

Rydberg atoms offer us another physical setup with Ising interactions that can be controlled by periodic driving fields. We consider a square lattice with one single Rydberg atom per site. An effective spin is formed with the states $| - \rangle_{\bf j} = | g \rangle_{\bf j}$, and 
$| + \rangle_{\bf j} = | r \rangle_{\bf j}$, corresponding to the ground and 
excited Rydberg state, respectively. 
The Hamiltonian describing this system is given by \cite{Weimer10natphys}
\begin{equation}
H_{\rm Ry} = 
\frac{U}{2} \sum_{<{\bf i} , {\bf j}>} \frac{\sigma^z_{\bf i} + 1}{2} \frac{ \sigma^z_{\bf j} + 1 }{2} + 
\frac{\Delta}{2} \sum_{\bf j} \sigma^z_{\bf j}.
\label{HRIsing}
\end{equation}
We have assumed fast decaying interactions between Rydberg states, such that the effective spin-spin interaction runs over first neighbours only. To obtain the compass model we need a periodic driving in the $x$-basis, since Ising interactions appear in the $z$-basis. Furthermore, we need to counteract the local longitudinal field given by $h^z = \Delta + U/2$. For this, we consider two counter-propagating driving fields inducing a Raman transition with wavectors ${\bf k}$ and $-{\bf k}$, and detunings $\omega_0 + \omega_{\rm d}$ and $\omega_0 - \omega_{\rm d}$, respectively.
\begin{eqnarray}
H_{\rm las}(t) &=& \nonumber \\ 
&& \sum_{\ii} \frac{\Omega_{\rm las}}{2} 
\left( e^{i  {\bf k}  {\bf r}_{\ii} - i (\omega_0 + \omega_{\rm d})t } \sigma^+ + {\rm H.c.}  \right) +
\nonumber \\ 
&& \sum_{\ii} \frac{\Omega_{\rm las}}{2} \left( e^{- i  {\bf k} {\bf r}_{\ii} - i (\omega_0 - \omega_{\rm d})t} \sigma^+ + {\rm H.c.}  \right).
\label{HRlaser}
\end{eqnarray}
We choose $\omega_0 = \Delta + U/2$ to counteract the effect of the longitudinal field. An additional coupling with vanishing effective wavevector is required to implement the term proportional to $\Omega$ in Eq. (\ref{pat}). Note that transitions between atomic ground and Rydberg states usually required two-photon Raman processes, such that the effective wavevector can vanish with a suitable orientation of the individual laser beams. We choose a term of the form,
\begin{equation}
H_{\rm \Omega}(t) = \frac{\Omega}{2} \sum_{\ii} \left( e^{-i \omega_0 t} \sigma^+ + {\rm H.c.} \right).
\label{HROmega}
\end{equation}
If we express the sum of contributions (\ref{HRIsing},\ref{HRlaser},\ref{HROmega}) in a frame rotating with the frequency $\omega)$, we get
\begin{eqnarray}
&&H_{\rm Ry} + H_{\rm las} + H_{\rm \Omega} \to \nonumber \\ 
&&\frac{U}{8} \sum_{<\ii,\jj>} \sigma^{z}_{\ii} \sigma^z_{\jj} + 
\frac{\Omega}{2}\sum_{\ii} \sigma^x_\ii +
\frac{\Omega_{\rm las}}{2} \sum_{\ii} \cos({\bf k} {\bf r}_\ii - \omega_{\rm d} t) \sigma^x_\ii.
\label{HRfinal}
\end{eqnarray}
After a rotation of the spin basis we obtain our targeted driven Ising model, with $J = - U/4$, and the phase gradient given by the laser optical phases, $\phi_\ii = d_0 ({\bf k} \ \ii)$.
\section{Conclussions}
We have shown that Ising interactions in a square lattice can be dressed by a periodic driving field and transformed into a quantum compass model. The key idea is to use site-dependent driving phases such that the dressed spin-spin interactions depend on the orientation of the bonds connecting lattice sites. 
We have also show that the ground state of the quantum compass model can be reached by adiabatically ramping down a field in the spin $z$-direction. By using symmetry arguments, we have found conditions under which one of the degenerate ground states can be reached, depending on the initial quantum state, and the number of sites in the lattice. 

We have discussed two possible implementations with ultracold bosonic atoms and Rydberg atoms in optical lattices. However, our ideas can be used in other experimental setups, for example in two-dimensional arrays of trapped ions. The latter system requires further investigation, since Ising interactions in trapped ion setups are long-ranged \cite{Porras04prl}, something that would lead to the implementation of long-range quantum compass models. Any experimental setup with spin interactions where couplings can be dressed by periodic fields is also amenable for the implementations of our ideas, like for example, arrays of superconducting qubits interacting with classical fields in transmission lines \cite{Porras12prl,Quijandria13prl}.
\section{Acknowlegments}
The research leading to these results has received funding from the People Programme (Marie Curie Actions) of the European Union's Seventh Framework Programme (FP7/2007-2013) under REA grant agreement no: PCIG14-GA-2013-630955. J.J. Garc{\'\i}a-Ripoll acknowledges funding by Mineco Project FIS2012-33022 and CAM Research Network QUITEMAD+.

\appendix
\section{\\Eigenvalues and eigenstates of the operator Z} \label{App:AppendixA}

In this section we find the eigenvalues and eigenstates of the operator $Z\equiv\prod_{\jj}\sigma^z_{\jj}$ corresponding to the Hilbert subspace spanned by the doubly-degenerate ground state of the quantum compass model. First, recall that the set of integrals of motion $\{P_{j_y}\}$ defined in Eq.\eqref{Prow} can be used to characterize the twofold ground state since they have different quantum numbers, namely either $(p_1,\ldots,p_{j_y},\ldots,p_m)=(1,\ldots,1,\ldots,1)$ or $(p_1,\ldots,p_{j_y},\ldots,p_m)=(-1,\ldots,-1,\ldots,-1)$. If we define the states corresponding to the previous eigenvalues as $|\pm\rangle_p$, this result can be written as $P_{j_y}|\pm\rangle_p=\pm|\pm\rangle_p$ $\forall j_{y}$. Alternatively, the set $\{Q_{j_x}\}$ can also be used for this purpose, giving the quantum numbers $q_{j_x}$. The corresponding eigenstates satisfy $Q_{j_x}|\pm\rangle_q=\pm|\pm\rangle_q$ $\forall j_{x}$, where the states $|\pm\rangle_q$ are in general different from $|\pm\rangle_p$. This result was proved in Ref.\cite{RefWorks:36} for a square lattice $l\times l$, and it can be straightforwardly generalized for a $n\times m$ lattice. 

Let us define the operators $X\equiv\prod_{\jj}\sigma^x_{\jj}$ and $Y\equiv\prod_{\jj}\sigma^y_{\jj}$. We can express $X$ and $Y$ in terms of the sets of operators $\{P_{j_y}\}$ and $\{Q_{j_x}\}$ as $X=\prod^n_{j_x}Q_{j_x}$ and $Y=\prod^m_{j_y}P_{j_y}$. Thus the states $|\pm\rangle_q$ and $|\pm\rangle_p$ are also eigenstates of $X$ and $Y$ respectively, leading to the eigenequations

\begin{align}
X|\pm\rangle_q&=(\pm 1)^n|\pm\rangle_q \\
Y|\pm\rangle_p&=(\pm)^m|\pm\rangle_p .
\end{align}

Notice, though, that the action of $X$ on the basis $|\pm\rangle_p$ is unknown, and similarly for $Y$ on the basis $|\pm\rangle_q$ . Bearing in mind the relation $\sigma^a\sigma^b=\mathbb{1}\delta^{ab}+i\sum\epsilon^{abc}\sigma^c$  ($a=1,2,3$) for the Pauli matrices, we find a useful equation to relate the previous operators, $XY=(i)^{m\times n}Z$. This allows us to calculate the commutators and anti-commutators for these operators, namely

\begin{align}
\{X,Y\}&=2 Re(i^{m\times n})Z \\
[X,Y]&=2 Im(i^{m\times n})Z.
\end{align}

We will show in the following how the above equations enable us to determine the eigenstates and eigenvalues of the operator Z in three different cases, depending on the parity of the lattice size, i.e. the parity of $n$ and $m$. Before doing so, let us prove a result that we shall require later. We would like to relate the basis  $|\pm\rangle_p$ and  $|\pm\rangle_q$, and this can be done by the fact that $\{P_{j_y},Q_{j_x}\}=0$ $\forall j_x,j_y$. Effectively, as $P_{j_y}|\pm\rangle_p=\pm|\pm\rangle_p$ $\forall j_{y}$, this implies $P_{j_y}(Q_{j_x}|\pm\rangle_p)=\mp(Q_{j_x}|\pm\rangle_p)$. Hence we infer that the state $Q_{j_x}|\pm\rangle_p$ has to be proportional to one of the eigenstates of $P_{j_y}$. The only possible solution in this particular case is $Q_{j_x}|\pm\rangle_p \propto |\mp\rangle_p$, since the choice $Q_{j_x}|\pm\rangle_p \propto |\pm\rangle_p$ leads to contradiction  $P_{j_y}|\pm\rangle_p=\mp|\pm\rangle_p$. Using an analogous argument for $P_{j_y}$, one can show that $P_{j_y}|\pm\rangle_q \propto |\mp\rangle_q$.

\subsection{Case 1: even n-even m}
In this case the previous relations take the following form,

\begin{align}
\{X,Y\}&=2Z \label{anticonm1}\\
[X,Z]&=[Y,Z]=0 \label{conm1} \\
X|\pm\rangle_q&=|\pm\rangle_q \label{eigenX1}\\
Y|\pm\rangle_p&=|\pm\rangle_p \label{eigenY1}.
\end{align}

Eq.\eqref{conm1} and Eq.\eqref{eigenY1} imply $Y(Z|\pm\rangle_p)=Z|\pm\rangle_p$. Hence the states $Z|\pm\rangle_p$ have to be proportional to the eigenstates of $Y$, either $Z|\pm\rangle_p \propto |\pm\rangle_p$ or $Z|\pm\rangle_p \propto |\mp\rangle_p$. The proper option can be inferred from  the anti-commutator relation \eqref{anticonm1}, which states that $Z=XY$. Effectively, introducing the state $|\pm\rangle_p$ to both sides of $Z=XY$, we find that $Z|\pm\rangle_p=X|\pm\rangle_p$. For even n, the operator $X$ can be expressed as an even product of $Q_{j_y}'s$ since $X=\prod^n_{j_x}Q_{j_x}$, each of these terms acting in a way that $Q_{j_x}|\pm\rangle_p \propto |\mp\rangle_p$ as proved before. Therefore, the only possibility is that $Z|\pm\rangle_p \propto |\pm\rangle_p$. The eigenvalues $z_{\pm}$ can also be calculated using Eq.\eqref{anticonm1}, since we have $_p\langle\pm|Z|\pm\rangle_p=_q\langle\pm|XY|\pm\rangle_p$, which leads to $_q\langle\pm|\pm\rangle_p z_{\pm}=_q\langle\pm|\pm\rangle_p$. Finally, we conclude that the states $|\pm\rangle_p$ are degenerate eigenstates of the operator Z with eigenvalues $z_{\pm}=1$. Similarly, the same result holds for the the states $|\pm\rangle_q$, i.e.  $Z|\pm\rangle_p= |\pm\rangle_p$ . 

\subsection{Case 2: even n-odd m}

The basic equations for this case can be written as follows,

\begin{align}
\{X,Y\}&=2Z(-1)^{(n/2 \times m)} \label{anticonm2}\\
[X,Z]&=[Y,Z]=0 \label{conm2} \\
X|\pm\rangle_q&=|\pm\rangle_q \label{eigenX2}\\
Y|\pm\rangle_p&=\pm|\pm\rangle_p \label{eigenY2}.
\end{align}

Equivalently to the previous case, as we have an even product of $Q_{j_y}'s$ we infer that $Z|\pm\rangle_p \propto |\pm\rangle_p$. Now using Eq.\eqref{anticonm2} the eigenvalues $z_{\pm}$ can be computed as $_q\langle\pm|\pm\rangle_p(-1)^{(n/2 \times m)} z_{\pm}= (\pm 1)_q\langle\pm|\pm\rangle_p $. Therefore, the operator $Z$ satisfies the relation $Z|\pm\rangle_p= \pm(-1)^{(n/2 \times m)}|\pm\rangle_p$. This case is trivially equivalent to the odd-even case by a proper rotation, and in such a case the operator $Z$ fulfills an analogous relation for the states $|\pm\rangle_q$.

\subsection{Case 3: odd n-odd m}

Finally, in this case our initial relations lead to a familiar set of equations,

\begin{align}
\{X,Y\}&=\{Y,Z\}=\{Z,X\}=0 \label{anticonm3}\\
[X,Y]&=2i^{m\times n}Z  \label{conm3} \\
[Y,Z]&=2i^{m\times n}X  \label{conm3} \\
[Z,X]&=2i^{m\times n}Y  \label{conm3} \\
X|\pm\rangle_q&=\pm|\pm\rangle_q \label{eigenX3}\\
Y|\pm\rangle_p&=\pm|\pm\rangle_p \label{eigenY3}.
\end{align}

Notice that these are essentially the same commutation and anti-commutation relations as the ones for the Pauli matrices, so we can use the well-known results of this representation to assure that the eigenvalues $Z$ in this subspace are $z_{\pm}=\pm 1$. One can also take advantange of the results for the Pauli matrices to express the eigenvectors of $Z$, $|\pm\rangle_z$ in terms of the eigenvectors of $X$ or $Y$. In particular, in term of the basis $|\pm\rangle_q$ we have  $|\pm\rangle_z=1/\sqrt{2}(|+\rangle_q+|-\rangle_q)$.
\bibliography{Compass}




\end{document}